\documentclass[aps,prl,twocolumn,showpacs,superscriptaddress,groupedaddress]{revtex4}
\usepackage{amsmath, amsthm, amssymb}
\usepackage{graphicx,epsfig}
\usepackage{color}
\usepackage{natbib}
\usepackage{setspace}

\begin{document}
\begin{spacing}{1.00}

\title{Shear thickening of highly viscous granular suspensions}

\affiliation{Department of Physics, University of Chicago, Chicago, IL 60637, USA}
\affiliation{James Franck Institute, University of Chicago, Chicago, IL 60637, USA}
\affiliation{Department of Mechanical Engineering and Materials Science, Yale University, New Haven, CT 06520, USA}

\author{Qin Xu} \affiliation{Department of Physics, University of Chicago, Chicago, IL 60637, USA} \affiliation{James Franck Institute, University of Chicago, Chicago, IL 60637, USA}
\author{Sayantan Majumdar} \affiliation{James Franck Institute, University of Chicago, Chicago, IL 60637, USA}
\author{Eric Brown} \affiliation{Department of Mechanical Engineering and Materials Science, Yale University, New Haven, CT 06520, USA}
\author{Heinrich M. Jaeger}  \affiliation{Department of Physics, University of Chicago, Chicago, IL 60637, USA} \affiliation{James Franck Institute, University of Chicago, Chicago, IL 60637, USA}

\date{\today}

\begin{abstract}
We experimentally investigate shear thickening in dense granular suspensions under oscillatory shear. Directly imaging the suspension-air interface, we observe dilation beyond a critical strain $\gamma_c$ and the end of shear thickening as the maximum confining stress is reached and the contact line moves. Analyzing the shear profile, we extract the viscosity contributions due to hydrodynamics $\eta_\mu$, dilation $\eta_c$ and sedimentation $\eta_g$. While $\eta_g$ governs the shear thinning regime,  $\eta_\mu$ and $\eta_c$ together determine the shear thickening behavior. As the suspending liquid's viscosity varies from 10 to 1000 cst, $\eta_\mu$ is found to compete with $\eta_c$ and soften the discontinuous nature of shear thickening.

\end{abstract}
\maketitle

Dense suspensions can increase their viscosity under rapid shear; i.e., they exhibit shear thickening (ST) [1-9]. To understand the origin of this ST transition, several mechanisms have been proposed. A hydro-cluster picture ascribes mild, continuous thickening to particle groups formed by viscous hydrodynamic interactions [10-12]. Dense granular suspensions can exhibit a much stronger shear thickening that can become discontinuous as a critical packing fraction is approached [13-15].  Recent works have related this to frictional particle interactions and dilation [2, 16, 17], similar to dry granular materials. In this scenario, the confining stress at the suspension-air interface keeps the suspension contained. Since granular systems prefer to dilate when made to flow, the normal stress and, subsequently, the friction between suspension and shear plate dramatically increases beyond a certain applied stress. Thus, the measured flow resistance shoots up as long as dilation is counteracted by confinement. 

In this granular mechanism, the frictional stress between solid particles is the dominant contributor to ST. The suspending liquid mainly acts as a boundary constraint to prevent expansion. Nevertheless, viscous hydrodynamic interactions, as another dissipation mechanism, still exist in the bulk [3, 7, 8], and lubrication and viscous drag could become significant when the suspending liquid is highly viscous. So far, however, an experimental characterization of how the hydrodynamics couples with dilation and affects the shear thickening of nearly jammed granular suspensions has been lacking.

 \begin{figure}[t]
\begin{center}
\includegraphics[width=70mm]{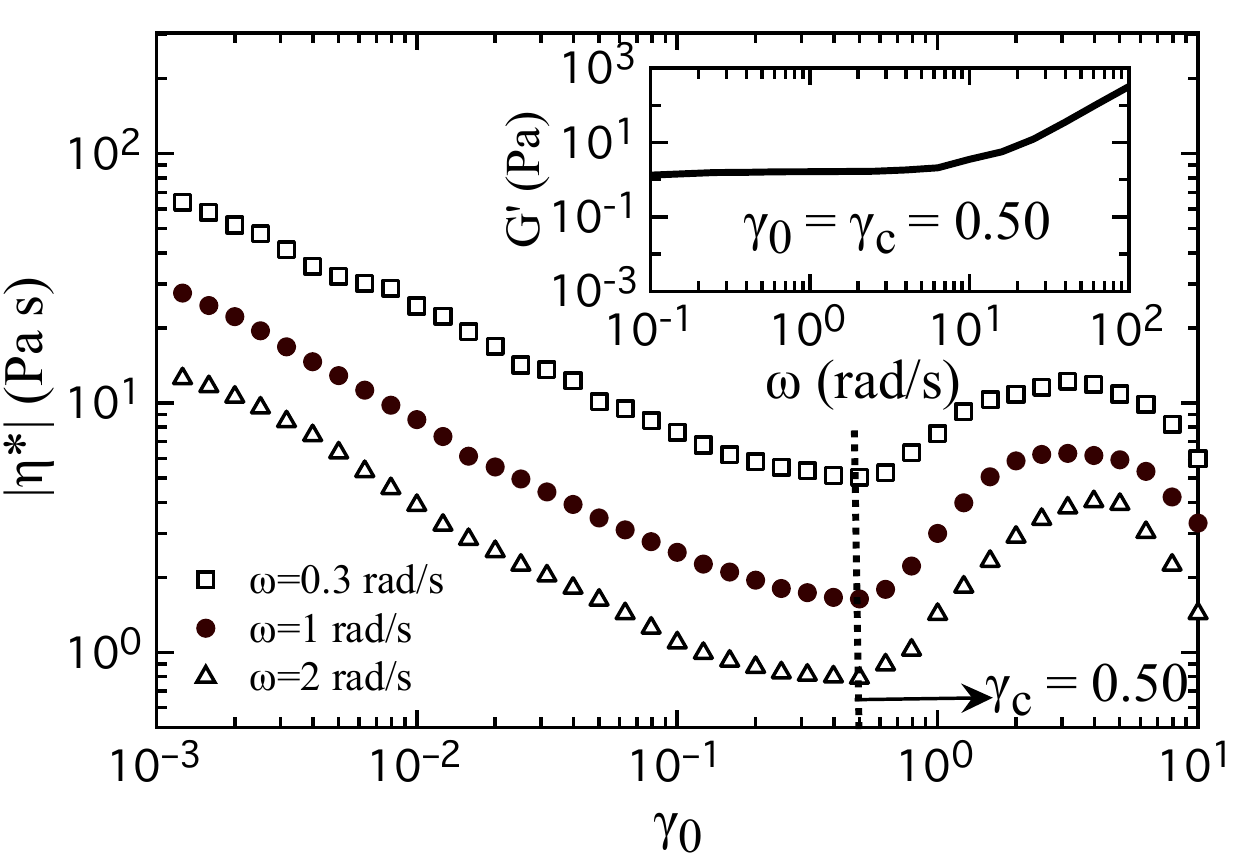}
\end{center}
	\caption{Rheological characterization of the viscosity $\vert \eta^{\star}\vert$ as a function of applied oscillatory strain amplitude $\gamma_0$ for $54\%$ ZrO$_2$ particles ($200 \mu$m) in 350 cst silicone oil at different oscillation frequencies $\omega$. Shear thickening starts at $\gamma_c \approx 0.50$ independent of $\omega$. Inset: storage modulus $G^{\prime}(\omega)$ vs. $\omega$ measured by fixing $\gamma_0$ at $\gamma_c $.}

\end{figure}

In this Letter, we address this issue by investigating dense granular suspensions across a wide range of suspending liquid viscosities. The suspended particles are chosen to be sedimenting so that the friction between particles can provide a known scaling for the onset stress of shear thickening [4]. To finely tune the relative displacement between particles, oscillatory shear with controlled amplitude is applied to the samples. Measuring both global rheology and local shear profile, we quantitatively extract the contributions from viscous hydrodynamics, confinement (``frustrated dilation") and sedimentation to the measured flow resistance.

 \paragraph{Experimental setup and protocol.---}
 Dense granular suspensions were prepared by adding ZrO$_2$ particles ($\rho_{\mbox{\small{ZrO}}_2}=3.92$ g/ml, $200\pm 10\:\mu$m) to silicone oils,  with packing fraction $\phi \approx54\%$. The particles were too large to exhibit measurable Brownian motion. Rheological measurements were performed in an Anton Paar rheometer with a 25mm diameter parallel-plate geometry. The gap size $d$ was varied from $1$ to $2$ mm. The top plate was set to apply a sinusoidal strain, $\gamma=\gamma_0 \sin (\omega t)$,  to the sample. We fit the measured shear stress to an oscillating waveform to obtain its the amplitude $\tau_0$ $[18-20]$. The magnitude of the complex viscosity is defined as $\vert \eta^{\star}\vert=\tau_0/\omega \gamma_0$,  the ratio of shear stress amplitude to applied strain rate. Before each measurement, the sample was pre-sheared to ensure repeatability. We used Vision Research Phantom V9 cameras with a macro lens (Nikon Micro 105 mm) to capture the dilation process. The frame rate was kept at 300 fps with spatial resolution $\sim12\: \mu$m/pixel. The samples were illuminated from the front by two white light sources (12 V/200 W, Dedolight).
 
\paragraph{Rheology.---}
We first focus on a suspension with oil viscosity of 350 cst. The rheology is quantified by ramping the strain amplitude $\gamma_0$ while keeping $\omega$ fixed in each round of measurement (Fig. 1). At the beginning of a ramp, $\vert \eta^{\star}\vert$ decreases with $\gamma_0$, that is, the samples shear thin. Beyond a critical strain $\gamma_c \approx  0.50$, $\vert \eta^{\star}\vert$ starts to increase with $\gamma_0$, indicating shear thickening. As previously observed by Fall \emph{et al.} [21], the value of $\gamma_c$ does not change with $\omega$. To rule out slip $[22]$, we performed the tests varying gap size and plate roughness (with a sand paper). No shift of $\gamma_c$ was observed. 

The onset of ST at $\gamma_c$ can be related to a stress scale. In oscillatory measurements, the onset stress is $\tau_{min}=\omega\gamma_c \vert \eta^{\star}\vert_{\gamma=\gamma_c}$. From each flow curve in Fig. 1, we calculate $\tau_{min}$ and find that the  suspension has $\tau_{min}\approx 0.75$ Pa, independent of $\omega$. At the same time, the elastic modulus $G^{\prime}(\omega)$ stays around $G^{\prime}_c\approx1.4$ Pa for $\omega$ from 0.1 to 10 rad/s (inset of Fig. 1(a)). Since the measured loss modulus is much smaller ($\sim 0.1$ Pa) in the same regime, the critical strain $\gamma_c$ is given by $\tau_{min}/G^{\prime}_0$ ($\approx$ 0.5). The same results were also found for suspensions made of different particles (see the Supplemental Material).  

\paragraph{Imaging of the interface and shear profile.---}
To quantify the particle dynamics and confirm the occurrence of dilation in the suspensions, we image the suspension-air interface while the strain is ramped up. The solid points in Fig. 2(a) show the rheological curve at $\omega=1$ rad/s. Figures (b)-(e) present an image sequence of the interface evolution. At $\gamma_0=0.002$, particles are completely contained in the liquid and no protrusions are observed at the surface (see Fig. 2(b)). As $\gamma_0$ approaches $\gamma_c \approx 0.50$, the shape of individual particles becomes visible (Fig. 2(c)). The local deformation of the interface can also be seen from the change of image brightness.

\begin{figure}[t]
\begin{center}
\includegraphics[width=85mm]{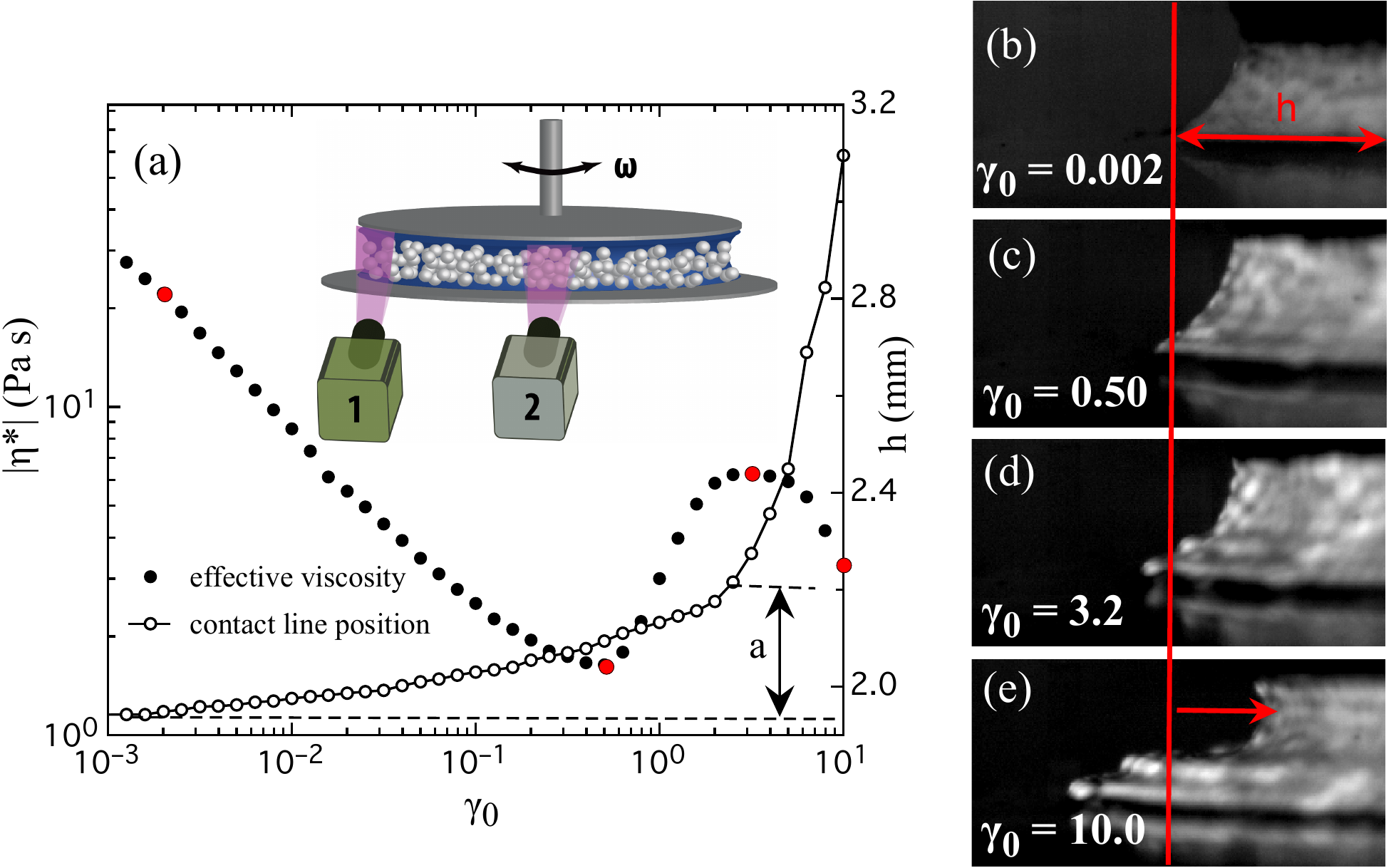}
\end{center}
\caption{(color online) Movement of the suspension-air interface. (a) Viscosity $\vert \eta^{\star}\vert$ and radial distance, $h$, to the outer edge of the suspension as a function of applied strain amplitude $\gamma_0$. Inset: Sketch of setup. Cameras 1 and 2 are used for imaging the interface and the velocity profile. (b)-(e) Images of the suspension boundary at specific $\gamma_0$ values, which are also displayed as red dots in (a). The vertical red line indicates the initial contact line position between suspension and substrate. }
\end{figure}

The vertical red line in Fig. 2 represents the initial position of the contact line between suspension and substrate. By tracking the outermost edge of suspension on the substrate, we plot the radial contact line position $h$ against $\gamma_0$ in Fig. 2(a) (hollow circles). At $\gamma_0=3.2$, the edge has moved out about one particle diameter $a$ and $\vert \eta^{\star} \vert$ starts to turn down (Fig. 2(d)). Thus, shear thickening stops when a full particle has been pushed out. This implies that the maximum confining stress from surface tension has been reached [16]. As a result, at higher strains or shear rates, the suspension thins, i.e. $\vert\eta^\star\vert$ decreases. With continuing increase of $\gamma_0$, the bottom portion of the suspension is further expanded (Fig. 2(e), $\gamma_0=10.0$). At the same time, the upper portion retracts (red arrow in Fig. 2(e)) [23]. 

To extract the shear profile, a second camera was placed right in front of the suspension. Figures 3(a) and (b) show typical images at $\gamma_0 = 1.0$ and $8.0$. The bottom layers in (b) are out of focus due to the expansion. From the recorded videos, the time averaged velocity fields are obtained by PIV (Particle Image Velocimetry). The resulting shear profile $v/v_p$ is plotted against $z/d$ for different $\gamma_0$ in Fig. 3(c), where  $v_p$ is the plate velocity and $z$ is the depth into the suspension, measured from the top. Before the onset of shear thickening, a shear band near the top plate is observed, extending a distance $d_s$. As $\gamma_0$ increases, the band gradually expands until spanning the entire gap ($d_s\sim d$ at $\gamma_0\gtrsim0.5$). With continuing increase of $\gamma_0$, particles near the bottom are pushed out and form effectively static layers that no longer participate in shear ($\gamma_0=3.2$ and $8.0$).
\begin{figure}[t]
\begin{center}
\includegraphics[width=75mm]{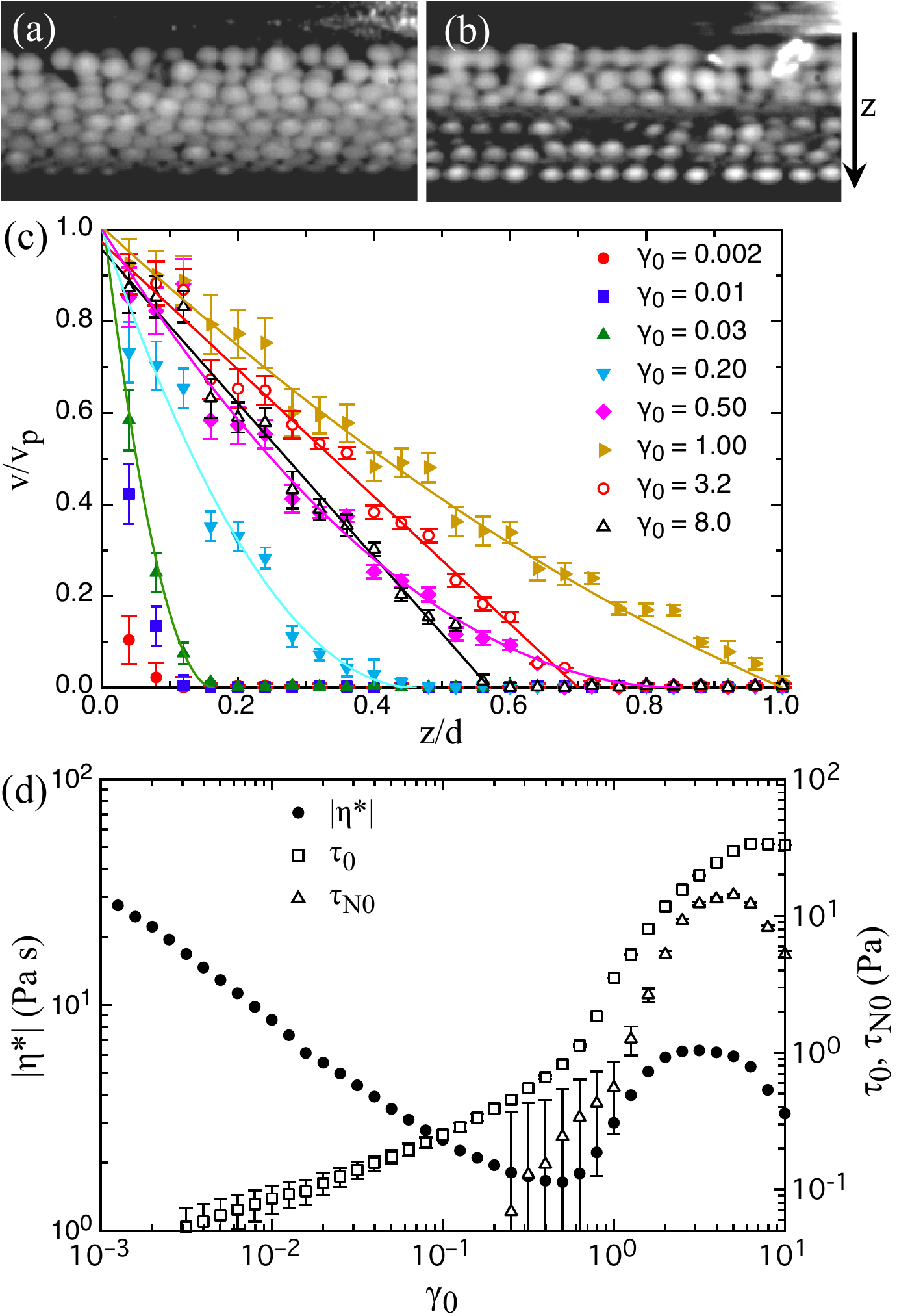}
\end{center}
\caption{(color online) Shear profile and stress scales. (a), (b) Front-view images of $54\%$ ZrO$_2$ ($200 \mu$m) in 350 cst silicone oil at $\gamma_0=1.0$ and $8.0$. (c) For the ZrO$_2$ suspension shown in Fig. 2, the time-averaged velocity profiles $v/v_p$, normalized by the speed of the top plate, are plotted against normalized depth $z/d$ for different  $\gamma_0$. Here $z=0$ corresponds to the top plate. The data points are obtained from PIV and the solid lines are fits to Eqs. (2)-(4). (d) Shear (hollow square) and normal stress (hollow triangle) amplitude together with the viscosity curves (solid circles).}
\end{figure}

\begin{figure*}[t]
\begin{center}
\includegraphics[width=150mm]{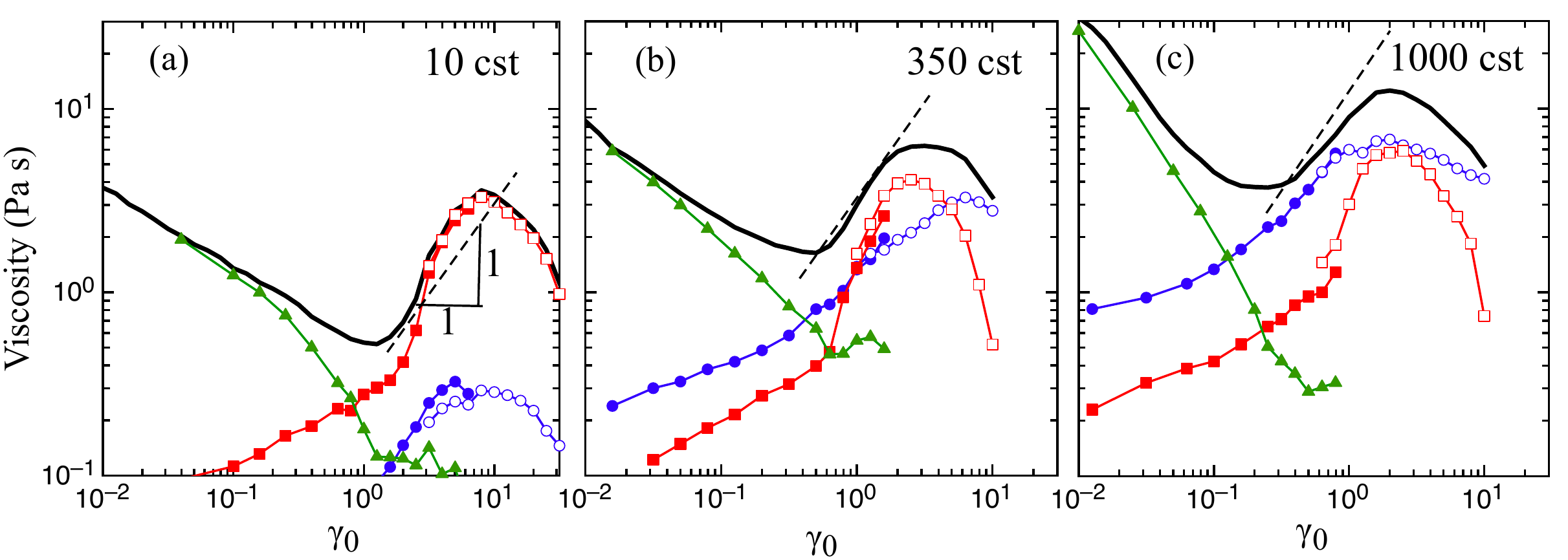}
\end{center}
\caption{(color online) The viscosity components due to hydrodynamics ($\eta_\mu$), dilation ($\eta_c$) and sedimentation ($\eta_g$) in the rheological  measurement. While the black lines is the measured viscosity $\vert\eta^{\star}\vert$ recorded by the rheometer, $\eta_\mu$, $\eta_c$ and $\eta_g$ are plotted in blue, red and green, respectively. The solid data points are obtained by the fit to Eqs. (2) and (3). The open symbols correspond to large-strain regime, where we directly calculate $\tau_c=\mu_c\tau_{N0}$ and $\eta_\mu=\vert\eta^{\star}\vert - \eta_c$. The suspending viscosities were varied to be (a) 10 cst, (b) 350 cst and (c) 1000 cst. The dashed lines indicate the Bagnoldian scaling: $\vert\eta^{\star}\vert\sim \gamma_0$.} 
\end{figure*}

In order to understand the evolution of the shear profile shown in Fig. 3(c) quantitatively, a microscopic constitutive relation has to be considered that accounts for the local stresses in the suspension. Quite generally, the shear stress $\tau_0$ contains contributions from hydrodynamic and inter-particle forces that could arise from a variety of sources [2, 10]. For dense granular suspensions of hard, non-Brownian particles, the dominant inter-particle forces arise from direct, frictional contact [2,3,8,16]. Thus, given a local, $z$-dependent shear rate $\dot{\gamma_l}$, 
\begin{equation}
\tau_0=\eta_\mu \dot{\gamma_l}+\tau_g z/d+\tau_c.
\end{equation}
Here, the first term represents the viscous hydrodynamic stress, contributing an amount $\eta_\mu$ to the measured overall viscosity, while the shear rate independent remainder reflects the inter-particle forces.  We have split this remainder into two parts to separate out the frictional stresses originating from sedimentation ($\tau_g$) and frustrated dilation ($\tau_c$) [16]. The second term in Eq. (1) scales linearly with $z$ due to gravity. To move the particles at the bottom layer ($z=d$), the required stress is at least $\tau_g$.  In a suspension of hard non-Brownian particles, $\tau_g = \mu_{c}\Delta \rho g d/ 15.3$, where $\mu_{c}$ is the friction coefficient ($\sim 0.8$ for ZrO$_2$), and gives the magnitude of the onset stress [16]. 

By integrating $\dot{\gamma_l}$ over $z$, we obtain the velocity profile: 
\begin{align}
&\frac{v}{v_p}=\frac{\tau_g}{2\omega\gamma_0\eta_{\mu}}(\frac{d_s-z}{d})^2 \:\:\:\:\:\:\:\:\:\:\:\:\:\:\:\:\:\:\:\:\:\:\:\:\: (d_s<d),\\
&\notag\frac{v}{v_p}=\frac{(\tau_0-\tau_g-\omega\gamma_0\eta_c)(d-z)}{\omega\gamma_0\eta_{\mu} d}+\frac{\tau_g}{2\omega\gamma_0\eta_{\mu}}(\frac{d-z}{d})^2 \\
&\:\:\:\:\:\:\:\:\:\:\:\:\:\:\:\:\:\:\:\:\:\:\:\:\:\:\:\:\:\:\:\:\:\:\:\:\:\:\:\:\:\:\:\:\:\:\:\:\:\:\:\:\:\:\:\:\:\:\:\:\:\:\:\:\:\:\:\:\:\:\:(d_s>d).
\end{align}
Here, $d_s= (\tau_0-\tau_c)d/\tau_g$ is the depth of the sheared layers and $\eta_c=\tau_c/(\omega\gamma_0)$ indicates the contribution to the viscosity caused by dilation against confining boundaries. Equation (2) thus corresponds to shear banding and Eq. (3) to fully developed shear flow within the effective gap region (which can become smaller than $d$ once the contact line moves out).

When $\gamma_0 \leq\gamma_c \approx 0.5$, the applied stress $\tau_0\leq 0.76$ Pa (see Fig.3(d)) is so small that $d_s<d$ and Eq.(2) predicts parabolic shear profiles. Beyond the thickening onset $(\gamma_0>\gamma_c\approx0.5)$, $\tau_0>\tau_g$ and therefore $d_s>d$, corresponding to global motion across the entire suspension, such that the shear profile is governed by Eq.(3) containing both linear and parabolic terms.

As $\gamma_0>2$, the sedimentation contribution ($\tau_g$) becomes negligible compared to the other terms in Eq. (1). Accordingly, Eq.(3) simplifies to a linear profile ${v}/{v_p}\approx{(\tau_0-\tau_c)z}/{\omega\gamma_0\eta_\mu d_{eff}}$, where $d_{eff}$ is the effective gap height, i.e., the depth at which $v(z) = 0$. In the absence of wall slip (see velocity profiles in Fig. 3(c)), the slope of $v/v_p$ plotted against $z/d_{eff}$ has to be unity. Writing $\tau_0$ and $\tau_c$ in terms of the associated viscosity contributions, this leads to

\begin{equation}
\eta_\mu+\eta_c=\vert\eta^{\star} \vert.
\end{equation}

In principle, the values of $\eta_\mu$ and $\eta_c$ can be obtained by fitting the profile data to Eqs.(2) \& (3), or for linear profiles, to Eq. (4). However, Eq. (4) is not sufficient to extract the values of both $\eta_\mu$ and $\eta_c$. To resolve this problem, we assume the normal stress in this regime is purely frictional such that $\tau_c\approx \mu_c\tau_{N0}$ [24], where $\tau_{N0}$ is the measured normal stress amplitude (see Fig.3 (d)), and $\eta_c=\mu_c\tau_{N0}/\omega\gamma_0$. Thus, in the large-strain regime, we calculate $\eta_\mu=\vert\eta^{\star}\vert-\mu_c\tau_{N0}/\omega\gamma_0$ from Eq.(4).

Based on the values of $\eta_\mu$ and $\eta_c$ at different strain amplitudes, we can assemble a diagram to indicate the contribution from different viscosity components to $\vert\eta^{\star}\vert$. Figure 4(b) sums up the results for the ZrO$_2$ beads in 350cst silicone oil. The black line shows $\vert\eta^{\star}\vert$ versus $\gamma_0$ trace from Fig. 1. On the same plot, the blue circles and red squares present the magnitudes of $\eta_\mu$ and $\eta_c$, respectively. While the solid points are from the fits to Eqs. (2) \& (3), open symbols correspond to the regime at large strain amplitudes when the shear profile is close to linear and the values of $\eta_\mu$ and $\eta_c$ are from Eq.(4) together with $\eta_c=\mu_c\tau_{N0}/\omega\gamma_0$. As we compare the two procedures within the same range of $1.0<\gamma_0<2.0$, the results from the fit and calculation consistently follow the same trend in the plot, suggesting that the assumption of frictional contacts and $\tau_c\approx\mu_c\tau_{N0}$ is reasonable. As shown in Fig. 4(b), the increase of $\eta_c$ is in general steeper than $\eta_\mu$. During the ST transition ($0.5<\gamma_0<4.0$), the ratio of $\eta_c$ and $\eta_\mu$ reaches a factor up to two, indicating the dilation is still a more important factor to shear thickening than the increase of hydrodynamic interactions.  

In addition, we define $\eta_g=\vert \eta^{\star}\vert-\eta_c-\eta_\mu$ as the viscosity component due to sedimentation. Considering Eq. (1), $\eta_g$ can be written as
\begin{equation}
\eta_{g}=\eta_\mu \left[ \Big\vert\frac{\operatorname{d}(v/v_p)}{\operatorname{d}(z/d)}\Big\vert _{z=0} -1\right].
\end{equation}
This term is determined by the slope of the shear profile near the top plate ($z=0$). In the presence of sedimentation, the shear profiles are nonlinear. For very small $\gamma_0$, the velocity gradient is significant at $z=0$ since the shear flow is localized within a very small layer of the suspension near the top plate. Therefore, $\eta_g$ is sufficiently large to be dominant in this regime (green triangles). As $\gamma_0$ increases, the flow region expands and the shear profile becomes less steep, such that $\eta_g$ decreases and the system shear thins until the dilation and hydrodynamic effects set in.

Panels (a) \& (c) in Fig. 4 show the behavior if the suspending oil viscosity is changed. The hear thinning regime is always dominated by $\eta_g$. Shear thickening, however, is determined by both $\eta_c$ and $\eta_\mu$. While  $\eta_c$ remains roughly the same, $\eta_\mu$ changes substantially when varying the suspending liquid viscosity from 10 to 1000 cst. For 10 cst oil (Fig. 4(a)), $\eta_c\gg\eta_\mu$, which explains why ST in this regime can be described by frustrated dilation alone [3, 7, 16]. For 1000 cst oil, on the other hand, $\eta_\mu$ rises about two orders of magnitude and we have $\eta_\mu>\eta_c$ for the entire measurement range. Thus, hydrodynamics starts to play an important role in controlling the ST behavior in the highly viscous limit.

Specifically, the increase in $\eta_\mu$  affects the steepness of the flow curve and softens the discontinuous nature of ST. For 10 cst oil (Fig. 4(a)), the slope in ST regime is steeper than the dashed lines, which present the classical Bagnoldian scaling, $\vert\eta^{\star}\vert\sim\gamma_0$ [6, 8]. While increasing the oil viscosity to 1000 cst (Fig. 4(c)), for instance, ST is found to be weaker than the Bagnoldian scaling since $\eta_\mu$ becomes crucial but increases in a way less steep than $\eta_c$.
   
 \paragraph{Conclusions.---} Under oscillatory shear, the ST onset in dense granular suspensions can be characterized by a critical strain $\gamma_c$ (Fig. 1) that signals the onset of dilation against the confining interface. ST sets in as particles begin to protrude through the interface while approaching $\gamma_c$ and stops when the contact line between suspension and substrate starts to move, reflecting that the maximum confining stress has been reached (Fig. 2). Modeling the shear flow by a local constitutive relation, we quantified the contributions from different sources to the measured viscosity (Figs. 3 \& 4). With increasing viscosity of the suspending liquid, the hydrodynamic contributions can become sufficiently large that it competes with the effects from frustrated dilation and softens the discontinuous nature of ST in granular suspensions. 
    
 \paragraph{Acknowledgements.---} We thank Ivo R. Peters and Carlos S. Orellana for many discussions. This work was supported by the National Science Foundation (NSF) MRSEC program under Grant No. DMR-0820054 and by the U.S. Army Research Office through Grant No. W911NF-12-1-0182. S. M. acknowledgs the support from a Kadanoff-Rice Postdoctoral Fellowship.

\end{spacing}

\begin{thebibliography}{99}
\bibitem{ST 1}
H. A. Barnes, J. Rheol \textbf{33}, 29 (1989).
\bibitem{ST 2}
E. Brown and H. M. Jaeger, Rep. Prog. Phys. \textbf{77}, 046602 (2014). 
\bibitem{ST 3}
R. Seto, R. Mari, J. F. Morris and M. M. Denn, Phys. Rev. Lett. \textbf{111}, 218301 (2013).
\bibitem{ST 4}
M. Wyart and M. E. Cates, Phys. Rev. Lett \textbf{112}, 098302 (2014).
\bibitem{ST 5}
M. E. Cates, J. P.Wittmer, J.-P. Bouchaud, and P. Claudin, Phys. Rev. Lett. \textbf{81}, 1841 (1998).
\bibitem{ST 6}
A. Fall, A. Lema\^{i}tre, F. Bertrand, D. Bonn and G. Ovarlez, Phys. Rev. Lett. \textbf{105}, 268303 (2010). 
\bibitem{ST 7}
C. Heussinger, Phys. Rev. E. \textbf{88}, 050201 (R) (2013).
\bibitem{ST 8}
N. Fernandez, R. Mani, D. Rinaldi, D. Kadau, M. Mosquet, H. L. Burger, J. C. Barrioz, H. J. Herrmann, N. D. Spencer, and L. Isa, Phys. Rev. Lett. \textbf{111}, 108301 (2013).
\bibitem{ST 9}
B. J. Maranzano and N. J. Wagner, J. Chem. Phys. \textbf{114}, 10 514 (2001).
\bibitem{hydrocluster_1}
N. J. Wagner and J. F. Brady, Physics Today \textbf{62}, 27 (2009).
\bibitem{hydrocluster_2}
J. F. Brady and G. Bossis, J. Fluid Mech \textbf{155}, 105-129 (1985).
\bibitem{hydrocluster_3}
X. Cheng, J. H. McCoy, J. N. Israelachvili, I. Cohen, Science \textbf{333}, 1276 (2011).
\bibitem{DST_1}
E. Brown, N. A. Forman, C. S. Orellana, H. Zhang, B. W. Maynor, D. E. Betts, J. M. DeSimone and H. M Jaeger, Nature. Mater. \textbf{9}, 220-224 (2010).
\bibitem{DST_2}
E. Brown and H. M. Jaeger, Phys. Rev. Lett. \textbf{103}, 086001 (2009).
\bibitem{DST_3}
C. B. Holmes, M. E. Cates, M. Fuchs, and P. Sollich, J. Rheol. \textbf{49}, 237 (2005).
\bibitem{Dilation_1}
E. Brown and H. M. Jaeger, J. Rheol.  \textbf{56}(4), 875-923 (2012).
\bibitem{Dilation_2}
A. Fall, N. Huang, F. Bertrand, G. Ovarlez and D. Bonn, Phys. Rev. Lett. \textbf{100}, 018301 (2008).
\bibitem{nonlinear_1}
M. Wilhelm, P. Reinheimer, M. Ortseifer, T. Neidh\"{o}fer and H. Spiess, Rheol. Acta \textbf{39}, 241-246 (2000).
\bibitem{nonlinear_2}
R. H. Ewoldt, A. E. Hosoi and G. H. McKinley, J. Rheol. \textbf{52}, 1427 (2008).
\bibitem{nonlinear_3}
N. Koumakis, A. Pamvouxoglou, A. S. Poulos and G. Petekidis, Soft Matter \textbf{8}, 4271 (2012).
\bibitem{critical strain}
A. Fall, F. Bertrand, G. Ovarlez and D. Bonn, J. Rheo. \textbf{56}, 575 (2012). 
\bibitem{slip boundary}
Y. S. Lee and N. J. Wagner, Rheol. Acta \textbf{42}, 199-208 (2003).
\bibitem{Stokes number}
Inertia is a negligible factor in this case since the Stokes number $St \approx 0.04 \ll 1$.
\bibitem{normal force}
P. Rognon, I.Einav and C. Gay, J. Fluid Mech. \textbf{689}, 77-96 (2011).

\end{thebibliography}
\end{document}